\documentstyle{elsart}
\input psfig

\begin{document}
\begin{frontmatter}
\title{Effect of the sample geometry on the second magnetization 
peak in single crystalline 
Ba$_{0.63}$K$_{0.37}$BiO$_3$ thick film}

\author[leipzig]{A. Yu. Galkin\thanksref{galkin}}, 
\author[leipzig]{Y. Kopelevich\thanksref{kope}},
\author[leipzig]{P. Esquinazi}
\author[leipzig]{A. Setzer}
\author[metal]{V. M. Pan} and 
\author[solids]{S. N. Barilo}
\address[leipzig]{Department of Superconductivity and Magnetism, 
Institut  f\"ur Experimentelle Physik II, 
Universit\"at Leipzig, Linn\'estr. 5, D-04103 
Leipzig, Germany}
\address[metal]{Institute of Metal Physics, National Academy of 
Sciences, Vernadsky str.36, 
Kiev 252041, Ukraine}
\address[solids]{Institute of Physics of Solids and Semiconductors, 
Academy of Science, Minsk 220072, 
Belarus}
\thanks[galkin]{On leave from Institute of Metal Physics, 
National Academy of Sciences, Kiev 252041, 
Vernadsky str.36, Ukraine. Supported by the S\"achsiches Staatsministerium
f\"ur Wissenschaft und Kunst}
\thanks[kope]{On leave from Instituto de Fisica, Unicamp, 
13083-970 Campinas, Sao Paulo, Brasil. Supported by the 
Deutsche 
Forschungsgemeinschaft under DFG IK 24/B1-1,
Project H, and by CAPES proc. No. 077/99.}

\begin{abstract}
Magnetization hysteresis loop $M(H)$ measurements performed on a 
single crystalline 
Ba$_{0.63}$K$_{0.37}$BiO$_3$ superconducting thick film reveal 
pronounced 
sample geometry dependence of the ``second magnetization peak" 
(SMP), i.e. a maximum in 
the width of $M(H)$ occurring at the field $H_{\rm SMP}(T)$. In 
particular, it is found that the SMP vanishes 
decreasing the film dimension. We argue that the observed sample 
geometry dependence of the SMP cannot be accounted for by models 
which 
assume a vortex pinning enhancement 
as the origin of  the SMP. Our results can be understood 
considering the 
thermomagnetic instability effect and/or non-uniform current 
distribution at $H < H_{\rm SMP}$ in large 
enough samples. 

\end{abstract}

\end{frontmatter}


Recently, the dependence of the second magnetization peak (SMP) - 
a maximum in the 
width of the magnetization hysteresis loop - on the sample 
geometry has been reported for Nb 
films \cite{1,2} and Bi$_2$Sr$_2$CaCu$_2$O$_8$ single crystals 
\cite{3}. In particular, it 
was found that the SMP 
vanishes in both superconductors for small enough dimensions of 
the samples. The occurrence of 
the SMP in large samples has been interpreted in terms of the 
thermomagnetic instability 
(TMI) effect that leads to a reduced, as compared to the 
isothermal critical state value, 
irreversible magnetization at $H < H_{\rm SMP}(T)$, where $H_{\rm 
SMP}(T)$ corresponds to the field  
at the SMP. On the other hand, Gurevich and Vinokur (GV) \cite{4} 
have suggested that the 
SMP can result from a non-uniform current flow at $H < H_{\rm 
SMP}(T)$ due to material 
inhomogeneities, without invoking TMI effects. This model \cite{4} 
also predicts the vanishing of  the 
SMP decreasing the sample size. Thus, the interpretations of SMP 
given in Refs.~\cite{1,2,3,4} are 
fundamentally different from those based on an increase of the 
critical current density, 
implying that a careful distinction between the different 
approaches is necessary. The study of 
the sample geometry dependence of  SMP is one of the experimental 
instruments allowing for 
such a distinction. 
\begin{figure}
\centerline{\psfig{file=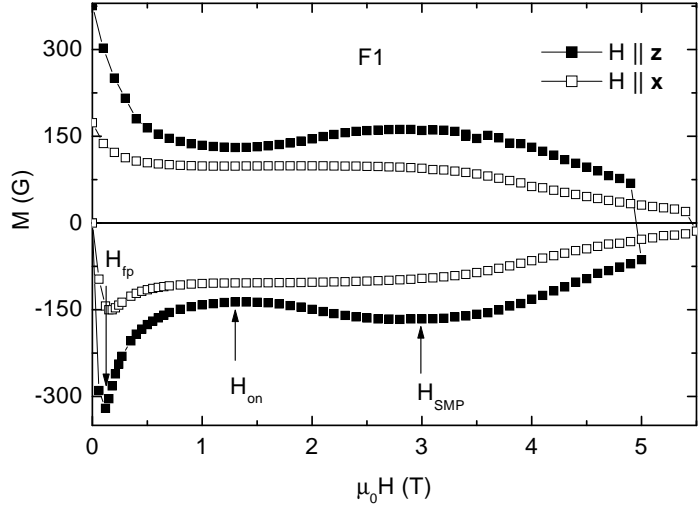,height=3.2in}}
\caption{Positive field region of magnetization hysteresis loops 
measured in the virgin film 
(sample F1) perpendicular ($H || z$) and parallel ($H || x$) to 
the main surface of the film. For $H || 
z$ configuration we indicate the magnetic fields corresponding to 
the first $H_{\rm fp}$ 
and second $H_{\rm SMP}$
magnetization peaks, as well as the SMP onset field $H_{\rm on}$.}
\end{figure}

The purpose of the present work is to verify the effect of the 
sample geometry on the SMP 
in the non-cuprate, isotropic high-$T_c$ ($T_c \simeq 30~$K) 
superconductor (Ba,K)BiO$_3$. 
The here studied sample is a Ba$_{0.63}$K$_{0.37}$BiO$_3$ single 
crystalline thick film grown on BaBiO$_3$ 
single crystal as a seed. The details of the film preparation are 
described elsewhere \cite{5}.  The 
good quality of the film was confirmed by  x-ray diffraction 
measurements (rocking 
curve FWHM  $\le 0.4^o$). The virgin film with dimensions $a 
\times b \times c = 2.9 \times 2.2 \times 0.25$~ mm$^3$,  
labeled as F1, was structured with the help of a thin wire in 10 
stripes of size $0.12 \times 2.2 \times 
0.25~$mm$^3$, i.e. with the cuts parallel to the sample b-
direction. This sample is labeled as F2. 
Then, the stripes size was further reduced to $120 \times 120 
\times 250~\mu$m$^3$. 
This sample, consisting of small-size bars, is labeled as F3. 

Isothermal 
magnetic hysteresis loop $m(H)$ measurements ($m(H)$ is the sample 
magnetic moment) were 
performed using a SQUID magnetometer (Quantum Design-MPMS7). The 
measurements 
were always made after cooling the sample in a zero applied 
magnetic field to the target 
temperature. In this work we restrict the presentation and 
discussion of the results obtained at $T = 5~$K.

Figure 1 shows magnetization curves $M(H) = m(H)/V$ ($V$ is the 
sample volume) 
measured with $H || x$ and $H || z $ (here and below $x, y$ and 
$z$ axes are directed along $a, b$ and $c$ 
dimensions of the virgin film, respectively). As seen in Fig.~1, 
the virgin magnetization $|M(H)|$
shows a local maximum at $\mu_0 H_{\rm fp} \simeq 0.1~$T, 
decreases with field and reaches a minimum at 
$\mu_0 H_{\rm on} \simeq 1.3~$T  and 
peaks again at $\mu_0 H_{\rm SMP} \simeq 3~$T (the SMP onset field 
$H_{\rm on}$ 
is nearly temperature independent at $T < 20~$K). 

\begin{figure}
\centerline{\psfig{file=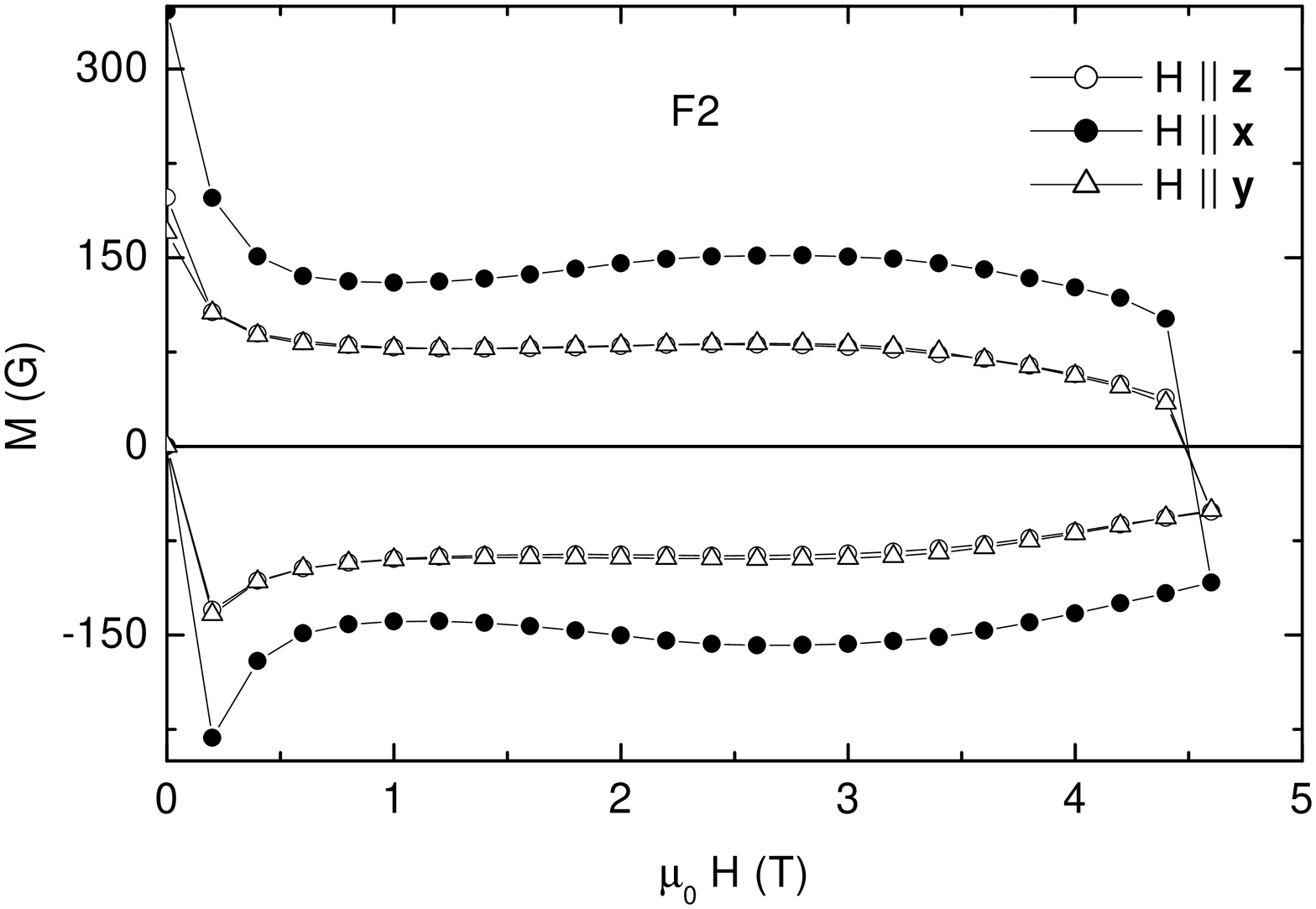,height=3.2in}}
\caption{Positive field region of magnetization hysteresis loops 
measured in the sample F2 for 
the three orthogonal field directions.}
\end{figure}
The first hint on the sample geometry dependence of the SMP comes 
from 
the pronounced difference in the hysteresis loops measured in the 
sample F1 with the field 
oriented perpendicular and parallel to the film$^,$s main surface. 
Indeed, as Fig. 1 demonstrates, 
the SMP is pronounced for the transverse $(H || z)$ geometry, 
whereas it is small for the 
parallel field configuration $(H || x)$. It has been shown, 
however, that the seeded growth-
induced anisotropy may affect the SMP \cite{6}. To check this we 
have measured $M(H)$ in the 
sample F2 for all three field orientations ($H || x, H || y,$ and 
$H || z$). Results of these 
measurements, given in Fig.~2, indicate that a well defined  SMP 
takes place for $H || x$, i. e. 
again for the transverse geometry, whereas for $H || y$ and $H || 
z$, the SMP is rather weak. This 
implies that the SMP is determined by the sample geometry, while 
the possible growth-induced 
anisotropy is irrelevant.

\begin{figure}
\centerline{\psfig{file=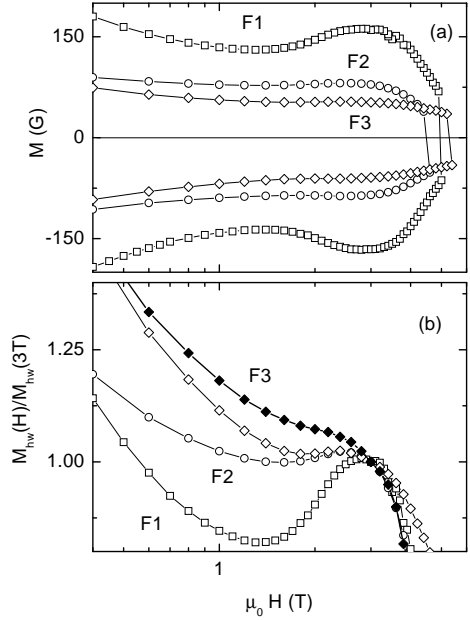,height=3.8in}}
\caption{ (a) Positive field region of magnetization hysteresis 
loops measured in the samples F1, 
F2, and F3 with $H || z$, and plotted in a semilogarithmic scale. 
(b) Half-width of the hysteresis 
loops $M_{\rm hw}(H)$ obtained for the samples F1, F2, F3 and 
normalized by the value at 
$\mu_0 H = 3~$T. $H || 
z$: open symbols, $H || x$: solid symbols (only for the sample 
F3).}
\end{figure} 
Figure~3(a) shows $M(H)$ measured in the samples F1, F2, and 
F3 with $H || z$, i.e. with the field direction parallel to the 
unchanged sample dimension $(c = 
250~\mu$m). Figure~3(b) shows the half-width of the hysteresis 
loops $M_{\rm hw}(H) = [M^+(H) - M^-(H)]/2$ obtained for the three 
samples and normalised by its value at $\mu_0 H = 3~$T ($M^+(H)$ 
and 
$M^-(H)$ are the magnetisation corresponding to the ascending and 
descending branches of the 
hysteresis loop). For completeness, $M_{\rm hw}(H)/M_{\rm hw}(3T)$ 
obtained for the sample F3 
with $H || x$ is also given in Fig.~3(b). As Fig.~3(a,b) 
demonstrates, the SMP is almost 
completely suppressed in the sample F3 which consists of the 
smallest-size bars. 

We stress, that the hysteresis loops measured in all samples and 
for different field 
orientations are symmetrical with respect to the zero 
magnetization axis. This indicates
that the hysteretic mechanism is dominated by bulk vortex 
properties as well as that the pinning induced irreversible 
magnetization $|M_{\rm irr}(H)|$ is much larger than the 
equilibrium magnetization $|M_{\rm eq}(H)|$. Thus, $|M_{\rm 
hw}(H)| (\sim |M_{\rm irr}(H)|)$
 is proportional to the critical 
current density $j_c(H)$ which can be calculated using the Bean 
model for either transverse or 
longitudinal geometry \cite{7}. We estimate for our samples $j_c 
\sim 10^9~$A/m$^2$ at $T = 5~$K, 
which slightly varies in 
the field interval $H_{\rm fp} < H < H_{\rm SMP}$. 

The magnetization time relaxation $M_{\rm irr}(t)$ measured in 
this work demonstrates that the vortex creep 
at 5 K only weakly affects $j_c$. The obtained logarithmic 
relaxation rate at 
$H < H_{\rm SMP},  \beta = - {\rm d}(\log (M_{\rm irr}))/{\rm 
d}(\log(t)) \sim 0.02$
 is in agreement with previous reports \cite{8}.
 
We start the discussion of the experimental results comparing the 
sample dimensions to the 
relevant vortex-pining-related characteristic lengths. 
At $H \sim H_{\rm on} \sim 1~$T~$ (<< H_{c2}(0))$ \cite{6} and 
taking the London penetration depth $\lambda (0) \sim 0.3~\mu$m, 
coherence length $\xi(0) \sim 4~$nm, the
Ginzburg-Landau parameter $\kappa =  \lambda/\xi  \sim 75$ 
\cite{6}, and $j_c \sim 
10^9~$A/m$^2$, we obtain:\\
\noindent --~the longitudinal collective 
pinning  length \cite{7} $L_c = (c_{44}/c_{66})^{1/2}R_c \sim 
3~\mu$m,\\
--~the transverse correlation radius $R_c = 8\pi r_p^2 
c_{44}^{1/2}c_{66}^{3/2}/W \sim 50~$nm 
(with the range of the pinning force $r_p \simeq \xi$),\\
--~the 
longitudinal size of the elastic cage $L_0 = (\epsilon_0/
c_{66})^{1/2} \sim 0.2~\mu$m, 
appearing in the theory of the 
disorder-driven transition between the flux-line-lattice (FLL) and 
the entangled vortex state 
\cite{9,10}.

Here $c_{44} \simeq H^2 \mu_0$ and $c_{66} \simeq \Phi_0 H 
/16\pi\lambda^2$
are the FLL tilt and shear moduli, 
respectively, $\epsilon_0 = (\Phi_0^2/4\pi\mu_0\lambda^2) 
\ln(\kappa)$ is the vortex line energy, $W = 
(F_pr_p^3c_{66}^2c_{44})^{1/2}/(1.5^{1/2}/32\pi^2)^{1/2}$ the 
parameter which 
measures the pinning strength \cite{7}, and $F_p = Bj_c$ the 
pinning force. Hence we conclude that the reduction of the sample 
dimension 
to  $\sim 100 \mu$m cannot affect the pinning efficiency of 
vortices. Thus, the observed sample 
size dependence of the SMP is clearly against an interpretation of 
the SMP based on an 
increase of $j_c(H)$. Note also that $L_c \gg L_0,$ i.e. vortices 
are in a different pinning regime from that discussed in 
Refs.~\cite{9,10}. 

Alternatively, Refs.~\cite{1,2,3,4} treat the SMP as resulting 
from either a TMI effect 
\cite{1,2,3} or from a non-uniform current flow due to material 
inhomogeneities \cite{4} 
which takes place at $H < H_{\rm SMP}$. 
According to the TMI theory, the condition for thermomagnetic 
stability is achieved for an 
effective sample dimension $s_{\rm eff} < s_{\rm crit} = 10H_{\rm 
fj}/4\pi j_c$ \cite{11}, 
where $H_{\rm fj}$ is the field at which the first 
magnetization jump (or reduction of the isothermal critical state 
magnetization which can 
occur in a rather smooth fashion even in Nb films \cite{1}) takes 
place. Within our interpretation (see 
also Refs.~\cite{1,2,3}), $H_{\rm fj} \simeq H_{\rm fp}$. Thus, 
taking the experimental values 
for $\mu_0 H_{\rm fp} \sim 0.1~$T and $j_c \sim 10^9~$A/m$^2$, we 
get $s_{\rm crit} \sim 100~\mu$m. 
For the sample F1 in the transverse geometry, $H || z, s_{\rm eff} 
\sim 
(bc/2)^{1/2} \sim 500~\mu$m \cite{12}, i.e. $s_{\rm eff} \gg 
s_{\rm crit}$. Therefore, a 
pronounced SMP  is expected as observed in this 
geometry  (see Fig. 1). 

Similarly, for $H || x$ the effective sample (F1) size 
\cite{11} $s_{\rm eff} = c/2 = 125~\mu$m, i.e. $s_{\rm eff} \sim 
s_{\rm crit}$. 
In this case, we would expect an essential suppression 
of the SMP as compared to the transverse geometry, which is also 
in agreement with 
the experimental observation (see Fig. 1). For an 
"intermediate" geometry $s_{\rm eff}$ cannot be easily estimated. 
For instance, the 
case of the sample F2 (long stripes with a nearly square cross 
section) with $H || x$, resembles 
that of a wire in a field applied normal to the wire axis, where 
the development of the SMP-like 
feature can be recognized as the length of the wire increases (see 
Fig. 3 in Ref. ~\cite{13}). 
Figure 3 (a, b) clearly demonstrates that the SMP ultimately 
vanishes as the sample 
dimensions become of the order of $\sim 100~\mu$m.
 
On the other hand, in Ref.~\cite{4} the hollow in the 
magnetization hysteresis loop in 
the field interval $H_{\rm fp} < H < H_{\rm SMP}$, has been 
attributed to a non-uniform current flow due to 
material inhomogeneties with a spatial scale $L \gg L_c$. This 
model predicts the vanishing of the 
SMP when the sample width becomes smaller than $w_c \sim L 
\sqrt{n}$, where $n = (1 + \beta)/\beta$. 
Taking $w_c \sim 100~\mu$m and the experimental value for $n \sim 
50~(\beta \sim 0.02)$ we get $L \sim
 14~\mu$m. Then the question arises: is the size of  the 
inhomogenities specific to our sample or it has a more general 
origin~? 

It is interesting to note that the SMP was essentially suppressed 
reducing the lateral sample size 
to $\sim 100~\mu$m also in Nb films \cite{2} and 
Bi$_2$Sr$_2$CaCu$_2$O$_8$ single crystals \cite{3},
 suggesting (within  the framework of the GV theory) a similar 
scale length $L$ in these otherwise different 
superconductors. It is tempting to associate this fact with the 
irregular fractal-type flux 
penetration observed at low fields in various superconductors such 
as, e. g., Nb films \cite{14}, 
Bi$_2$Sr$_2$CaCu$_2$O$_8$ and YBa$_2$Cu$_3$O$_{7-x}$ single 
crystals \cite{15}, 
Tl$_2$Ba$_2$CuO$_{6+x}$ and YBa$_2$Cu$_3$O$_{7-x}$ 
thin films \cite{16}, and assume $L$ as a characteristic width of 
flux droplets or fingers of size
$\sim 10 \ldots 100~\mu$m \cite{14,15,16}. If such a scenario is 
correct, a combined effect of a non-uniform current 
flow due to the fractal-type flux penetration and TMI can be 
expected.

We note further that local heating leading to avalanche-like flux 
penetration in Tl$_2$Ba$_2$CuO$_{6+x}$ and YBa$_2$Cu$_3$O$_{7-x}$ 
thin films has been proposed in Ref.~\cite{16}. The dependence of 
the fingering on thermal 
conditions as well as on the sample size was also reported for 
untwined YBa$_2$Cu$_3$O$_{7-x}$
 single crystal \cite{17}. Further theoretical and experimental 
work is needed to clarify 
the joint contributions of the TMI and non-uniform current flow to 
the development of the SMP 
in numerous superconductors as well as in (Ba,K)BiO$_3$ \cite{18}.

To conclude, we reported the vanishing of the SMP decreasing the 
sample size 
in isotropic non-cuprate high-$T_c$ superconductor 
Ba$_{0.63}$K$_{0.37}$BiO$_3$. The results 
provide evidence that the SMP is not related to the enhancement of 
the pinning efficiency of 
vortices and strongly suggest that the SMP originates from a 
thermomagnetic instability effect 
and/or from a non-uniform current flow. 

\ack
The authors thank F. Mrowka and A. V. Pan for assistance. This work is partially 
supported by the German Israeli Foundation for
Scientific Research and Development under G-553-191.14/97.

\end{document}